\documentclass[journal]{IEEEtran}
\usepackage{cite}
\usepackage{amsmath,amssymb,amsfonts}
\usepackage{graphicx}
\usepackage{textcomp}
\usepackage{algpseudocode}
\usepackage{multirow}
\usepackage{epstopdf}
\usepackage{amsmath}
\usepackage{lipsum} 
\usepackage{float} 
\usepackage{subcaption}
\usepackage[utf8]{inputenc}
\usepackage{mdframed}
\usepackage[ruled,boxed]{algorithm2e}
\usepackage{xcolor}

\definecolor{bottomcolor}{RGB}{247, 248, 236}

\begin{document}
\title{\huge SCDM: Unified Representation Learning for EEG-to-fNIRS Cross-Modal Generation in MI-BCIs}
\author{Yisheng Li and Shuqiang Wang, \IEEEmembership{Senior Member, IEEE} }

\maketitle

\begin{abstract}
Hybrid motor imagery brain-computer interfaces (MI-BCIs), which integrate both electroencephalography (EEG) and functional near-infrared spectroscopy (fNIRS) signals, outperform those based solely on EEG. However, simultaneously recording EEG and fNIRS signals is highly challenging due to the difficulty of colocating both types of sensors on the same scalp surface. This physical constraint complicates the acquisition of high-quality hybrid signals, thereby limiting the widespread application of hybrid MI-BCIs. To facilitate the acquisition of hybrid EEG-fNIRS signals, this study proposes the spatio-temporal controlled diffusion model (SCDM) as a framework for cross-modal generation from EEG to fNIRS. The model utilizes two core modules, the spatial cross-modal generation (SCG) module and the multi-scale temporal representation (MTR) module, which adaptively learn the respective latent temporal and spatial representations of both signals in a unified representation space. The SCG module further maps EEG representations to fNIRS representations by leveraging their spatial relationships. Experimental results show high similarity between synthetic and real fNIRS signals. The joint classification performance of EEG and synthetic fNIRS signals is comparable to or even better than that of EEG with real fNIRS signals. Furthermore, the synthetic signals exhibit similar spatio-temporal features to real signals while preserving spatial relationships with EEG signals. Experimental results suggest that the SCDM may represent a promising paradigm for the acquisition of hybrid EEG-fNIRS signals in MI-BCI systems.

\end{abstract}


\section{Introduction}
\label{sec:introduction}
\IEEEPARstart{T}{he} motor imagery brain-computer interface (MI-BCI) technology assists individuals with motor impairments by decoding neural signals from the central nervous system during imagined limb movements and then translating them into commands to control external devices \cite{bci}. A primary focus in MI-BCI research is the classification of left-right motor imagery (MI). Electroencephalography (EEG), which measures electrical activity on the cortical surface by placing electrodes on the scalp, is the most widely used technique for recording central nervous system signals in MI-BCI applications due to its high temporal resolution and quick response to stimuli \cite{hbci}.

Despite its advantages, EEG suffers from low spatial resolution and susceptibility to artifacts, which limit MI decoding effectiveness \cite{eeg_1, eeg_2}. In contrast, functional near-infrared spectroscopy (fNIRS) offers higher spatial resolution and greater resistance to artifacts. Moreover, it measures hemodynamic responses instead of electrical reactions in the brain, providing valuable neurophysiological insights for MI tasks \cite{fnirs_1, fnirs_2}. Combining EEG with fNIRS to construct multimodal MI decoding models, the hybrid MI-BCI system extracts discriminative features from each modality and integrates their information during pattern classification. Growing evidence indicates that hybrid MI-BCIs based on EEG-fNIRS outperform MI-BCIs based solely on EEG in terms of classification accuracy and stability \cite{eeg_fnirs_1, eeg_fnirs_2, eeg_fnirs_3, eeg_fnirs_4, eeg_fnirs_5}. These studies utilize independent EEG and fNIRS signals, as well as combined EEG-fNIRS signals, as inputs for MI classification tasks. Results show that hybrid signals significantly enhance classifier performance, surpassing that of independent EEG signals alone.

However, it is highly challenging to co-locate EEG and fNIRS sensors on the same scalp surface, primarily due to the consideration of the impact of dense hair interference on light signals \cite{hbci_hair_1, hbci_hair_2}, the influence of fNIRS source-detector distance \cite{hbci_sensor_1, hbci_sensor_2}, and the problem of matching EEG electrodes with fNIRS signal recording locations \cite{hbci_overlap}. This physical constraint hinders the acquisition of multi-channel, high-spatial-resolution fNIRS signals that overlap with EEG channel locations. As a result, in practical research, different device layouts must be designed based on various experimental paradigms to balance signal quality and device arrangement. Additionally, the complexity of pre-experimental preparations makes it inconvenient to collect joint signals frequently. These severely undermine the applicability of hybrid MI-BCI.

With the rapid advancement of artificial intelligence technologies, generative models have opened up possibilities for cross-modal generation of brain functional data \cite{ai_1, ai_2}. A prominent approach is the generative adversarial network (GAN) \cite{gan}, initially used for image generation, then extended to medical image computing \cite{gan_detection_1, gan_detection_2, gan_detection_3}, and later successfully extended to cross-modal generation in medical imaging \cite{gan_generation_1, gan_generation_2, gan_generation_3, gan_generation_4, gan_generation_5}. For example, Ben-Cohen et al. \cite{gan_generation_1} combined a fully convolutional network with conditional GAN to generate simulated positron emission tomography (PET) data from given computerized tomography (CT) inputs, providing a solution to reduce the need for expensive and radioactive PET/CT scans. Yang et al. \cite{gan_generation_2} utilized Structure-Constrained CycleGAN for unsupervised synthesis from MRI to CT. Dar et al. \cite{gan_generation_3} employed conditional GAN to synthesize images with different contrasts from a single MRI modality, enhancing diagnostic information diversity in MRI.

GANs, while implicitly representing image distributions, may nonetheless suffer from limited sample fidelity, thereby restricting the quality and diversity of synthesized images \cite{gan_vs_ddpm}. Additionally, training instability and mode collapse present major challenges to GAN. Recently, a novel generative model, the diffusion model \cite{ddpm, score_dm, ddim, latent_dm}, which relies on explicit likelihood representation and progressive sampling, has emerged. The diffusion model, inspired by the variational iteration method \cite{vim_1, vim_2}, generates data by progressively incorporating and removing noise, offering more stable training and enabling diverse sample generation, with demonstrated potential for cross-modal generation of medical images. For example, Jiang et al. \cite{dm_1} proposed the first diffusion-based multimodal MRI synthesis model, namely, the Conditional Latent Diffusion Model CoLa-Diff. Muzaffer et al. \cite{dm_2} introduced SynDiff, a novel approach based on adversarial diffusion modeling, achieving efficient and high-fidelity modality transformations and superior performance in multi-contrast MRI and MRI-CT synthesis. Li et al. \cite{dm_3} presented FGDM, which utilizes frequency-domain filters to guide the diffusion model for structure-preserving image transformations, successfully accomplishing tasks such as cone-beam CT-to-CT transformations across different anatomical sites and cross-institutional MR imaging conversions. 

Generative models have made notable progress in the cross-modal generation of brain imaging. However, there remains substantial exploration space, particularly in the realm of 1-dimensional time series data, such as the cross-modal generation of fNIRS signals from EEG, which warrants further investigation. 

This study introduces the spatio-temporal controlled diffusion model (SCDM), which utilizes EEG signals to cross-modally generate fNIRS signals. The synthetic fNIRS signals approach the qualities of real signals and can substitute real fNIRS to achieve comparable or even superior classification performance in MI tasks. The contributions of this paper include:

\begin{itemize}
    \item[(i)] A new paradigm for hybrid EEG-fNIRS signal acquisition in MI-BCI is proposed. By circumventing the constraints of simultaneous EEG-fNIRS signal recording, the proposed method allows for one-time fNIRS signal acquisition for long-term reuse. To our knowledge, this is the first study to achieve cross-modal generation from EEG to fNIRS signals.
    \item[(ii)] A spatial cross-modal generation (SCG) module is developed, integrating an improved time-series-based 2-dimensional self-attention mechanism. This module not only effectively learns the latent spatial representations of both signals but also achieves accurate representation mapping from EEG to fNIRS, preserving the spatial correspondence of fNIRS and EEG signals.
    \item[(iii)] A multi-scale temporal representation (MTR) module is designed, combining causal dilation convolution with depth-wise separable convolution to capture diverse temporal representations and eliminate spatial information interference. These convolutions control the learning of spatial and temporal representations as two independent processes, enhancing the accuracy of representation.
\end{itemize}

\section{Method}

\subsection{Diffusion Model}

In the diffusion model, the diffusion process at each time step $t$ involves adding noise $\epsilon_{t}$ sampled from a Gaussian distribution to the fNIRS time series $f_t$, where $t$ ranges from $1$ to $T$. After $T$ steps, the original sequence $f_0$ is transformed into noisy data $f_T$ that conforms to a standard Gaussian distribution. The diffusion model incorporates a U-Net architecture, which takes $f_t$ as input and outputs the predicted noise $\epsilon_{\theta, t}$ as an estimate of $\epsilon_{t}$, where $\theta$ represents the parameters of the U-Net. The optimization objective of the diffusion model is to minimize the $L_2$ loss between these two types of noise. Once trained, the diffusion model uses randomly sampled Gaussian noise as $\hat{f}_T$ and then gradually removes noise $\epsilon_{\theta, t}$ via backpropagation to obtain $\hat{f}_0$, i.e., the synthetic fNIRS time series.

\subsection{SCDM}

\begin{figure*}[!t]
\centering
\includegraphics[width=\linewidth]{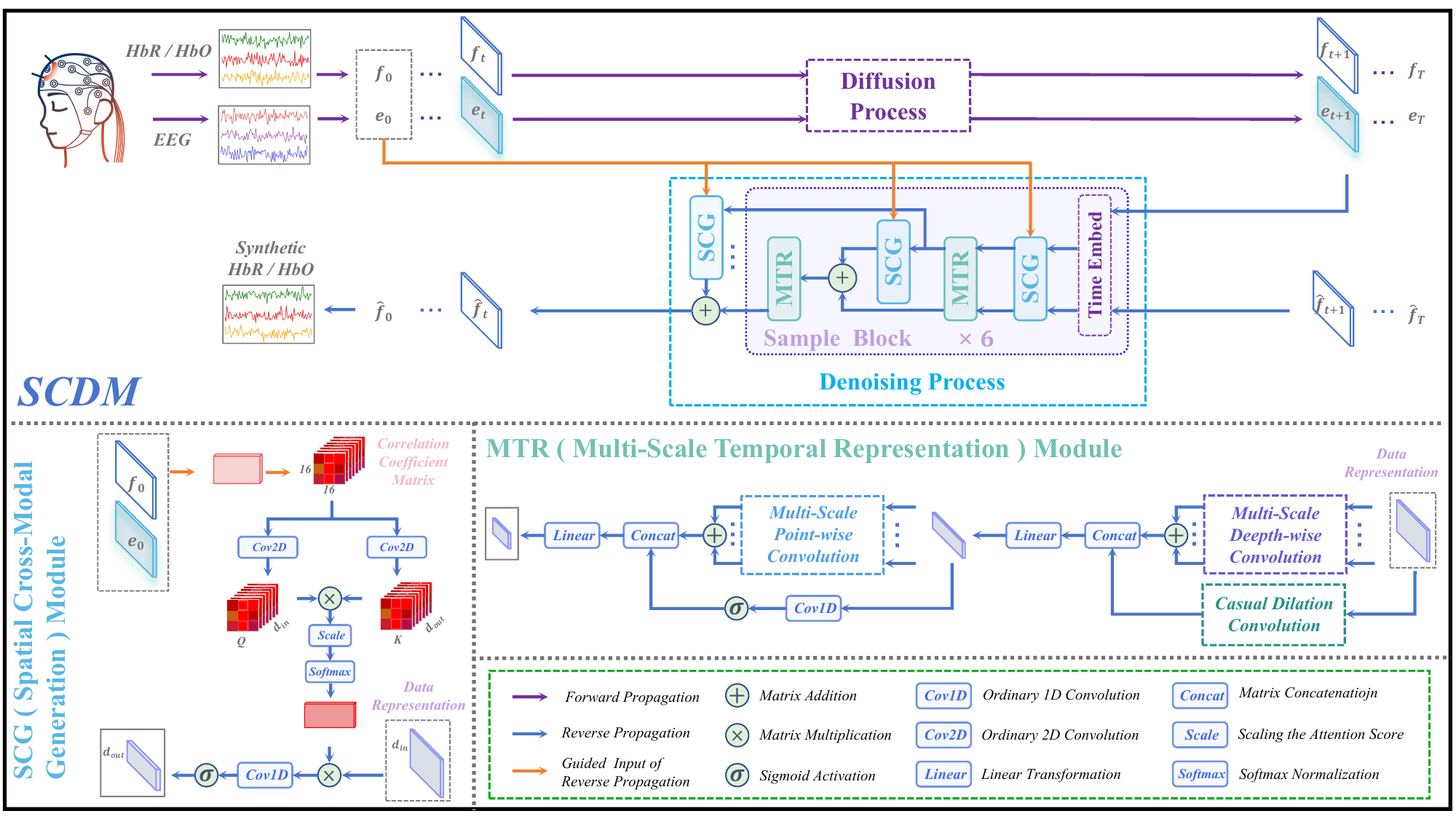}
\caption{Framework diagram of SCDM, and its core modules SCG and MTR.}
\label{architecture}
\end{figure*}

The proposed SCDM, illustrated in Fig. \ref{architecture}, distinguishes itself from the vanilla diffusion model by utilizing time series data from both EEG and fNIRS modalities to synthesize an fNIRS sequence. There is a certain degree of coupling between fNIRS and EEG signals, as hemodynamic changes are typically induced by neural activity. Based on this observation, one might assume that fNIRS and EEG sequences share similar distributions. While this assumption isn't universally applicable, the study introduces a method involving the selection of forward propagation parameters, especially time step numbers $T$ and noise scales $\beta_1,\beta_2,...,\beta_T$, to ensure that both $f_T$ and $e_T$ conform closely to Gaussian distributions with minimal mutual discrepancies. This approach forms a reasonable basis for considering $e_0$ and $f_0$ to have approximately similar distributions.

The optimization objective aims to minimize the pairwise distribution discrepancies $W$ between the fNIRS noise $f_T$ and EEG noise $e_T$ after forward propagation and the randomly sampled fNIRS noise $\hat{f}_T$ during backward propagation. The combined discrepancy $W$ is formulated as:

\begin{align}
W = W(f_T, \hat{f}_0) + W(e_T, \hat{f}_0) + W(f_T, e_T)
\end{align}
Here, the Wasserstein distance $W(P, Q)$ quantifies the actual differences between distributions $P$ and $Q$. Parameter selection precedes model training. Once suitable forward propagation parameters are identified, the original sequences $e_0$ and $f_0$ can undergo an identical forward propagation process.

The U-Net of SCDM takes the EEG and fNIRS time series at time step $t$, denoted as $e_t$ and $\hat{f}_t$ respectively, as input and predicts the noise $\epsilon_{\theta,t}$. Each $\hat{f}_t$ is derived from the backward-sampled output of the previous step $t+1$, while $e_t$ is obtained from the forward-propagated output of the current step $t$. The U-Net consists of 6 sample blocks. The first 3 blocks perform downsampling, halving the channel count and sequence length of $e_t$ and $f_t$, while the last 3 blocks perform upsampling, doubling those of $e_t$ and $f_t$. Each sample block primarily comprises the SCG and MTR two core modules, with the Time Embed module first embedding the time step information. The structure of all sample blocks is nearly identical, with the only difference being that the MTR module in the upsampling blocks uses transposed convolutions or bilinear interpolations instead of standard convolutions. To facilitate the process of the two time series within the sample blocks, prior to inputting them into the U-Net, convolution and linear transformation operations are applied to convert $e_t \in \mathbb{R}^{N\times 30 \times 4000}$ and $f_t \in \mathbb{R}^{N\times 36 \times 256}$, which have 30 channels and 36 channels respectively, into the same shape of $N\times 32\times 256$, where $N$ denotes the batch size.

\begin{algorithm}
\caption{Training Phase of the SCDM}
\begin{mdframed}[backgroundcolor=bottomcolor,rightline=false,leftline=false,topline=false,bottomline=false,innerleftmargin=5pt,innerrightmargin=5pt,userdefinedwidth=0.96\linewidth,innerbottommargin=5pt,innertopmargin=5pt]

\textbf{Define:}

$T$: time step numbers

$t$: time step which ranges from $1$ to $T$

$e_t$: EEG data in the diffusion process at time step $t$

$f_t$: fNIRS data in the diffusion process at time step $t$

$\hat{f}_t$: fNIRS data in the denoising process at time step $t$

$\beta_t$: noise scale at time step $t$

$\epsilon_t$: noise in the diffusion process at time step $t$

$\epsilon_{\theta}$: the U-Net for noise prediction

$\hat{\epsilon}_{\theta, t}$: predicted noise by the U-Net at time step $t$

$\mathcal{L}(\theta)$: loss function

$W$: distribution discrepancy

\BlankLine

\textbf{Input:}

$e_0$: original EEG data

$f_0$: original fNIRS data

\BlankLine

\textbf{Diffusion Process:}

$W \gets W(f_T, \hat{f}_T) + W(e_T, \hat{f}_T) + W(f_T, e_T)$

determine $T$ and {$\beta_1$, $\beta_2$, ..., $\beta_T$} by minimizing $W$

$\alpha_t \gets 1-\beta_t$

\For{$t \gets 1$ \KwTo $T$}
{
    
    Sample $\epsilon_t \sim \mathcal{N}(0, I)$ 
    
    $e_t \gets \sqrt{\alpha_t} e_{t-1} + \sqrt{\beta_t} \epsilon_t$
    
    $f_t \gets \sqrt{\alpha_t} f_{t-1} + \sqrt{\beta_t} \epsilon_t$
}

\BlankLine

\textbf{Denoising Process:}

Sample $\hat{f}_T \sim \mathcal{N}(0, I)$

\For{$t \gets T$ \KwTo $1$}
{
    
    $\hat{\epsilon}_{\theta, t} \gets \epsilon_{\theta}(e_t, \hat{f}_t, e_0, f_0)$ 
    
    $\hat{f}_{t-1} \gets \frac{1}{\sqrt{\alpha_t}} \left( f_t - \frac{\beta_t}{\sqrt{1 - \bar{\alpha}_t}} \hat{\epsilon}_{\theta, t} \right)$
}
\BlankLine

\textbf{Backpropagation:}

$\mathcal{L}(\theta) = {\left \| \epsilon_t - \hat{\epsilon}_{\theta, t} \right \|}^2_2$ 

update model parameters by minimizing $\mathcal{L}(\theta)$

\BlankLine

\end{mdframed}
\end{algorithm}

\begin{algorithm}

\caption{Inference Phase of the SCDM}
\begin{mdframed}[backgroundcolor=bottomcolor,rightline=false,leftline=false,topline=false,bottomline=false,innerleftmargin=5pt,innerrightmargin=5pt,userdefinedwidth=0.96\linewidth,innerbottommargin=5pt,innertopmargin=5pt]
\textbf{Define:}

$T$: time step numbers

$t$: time step which ranges from $1$ to $T$

$\hat{f}_t$: fNIRS data in the denoising process at time step $t$

$\epsilon_{\theta}$: the U-Net for noise prediction

$\hat{\epsilon}_{\theta, t}$: predicted noise by the U-Net at time step $t$

\BlankLine

\textbf{Input:}

$e_t$: EEG data in the diffusion process at time step $t$

$f_0$: original fNIRS data

$\beta_t$: Noise scale at time step $t$

\BlankLine

\textbf{Inference Process:}

Sample $\hat{f}_T \sim \mathcal{N}(0, I)$ 

$\alpha_t \gets 1-\beta_t$

\For{$t \gets T$  \KwTo $1$}
{
    
    $\hat{\epsilon}_{\theta, t} \gets \epsilon_{\theta}(e_t, \hat{f}_t, e_0, f_0)$ 
    
    $f_{t-1} \gets \frac{1}{\sqrt{\alpha_t}} \left( f_t - \frac{\beta_t}{\sqrt{1 - \bar{\alpha}_t}} \hat{\epsilon}_{\theta, t} \right)$
}

\BlankLine

\textbf{Output:}

Synthetic fNIRS signal $\hat{f}_0$

\BlankLine

\end{mdframed}
\end{algorithm}

\subsection{SCG Module}

The SCG module halves or doubles the time series channels without altering the sequence length, supporting various input and output combinations.

\subsubsection{Representation Mapping}
When the inputs include the original sequences $e_0$ and $f_0$ along with the EEG representation, the SCG module maps the EEG representation to the fNIRS representation.

Firstly, $e_0$ and $f_0$ are used to compute the distance correlation coefficient matrices between the $30$ EEG channels and $36$ fNIRS channels. Due to the differing lengths of the two sequences, Pearson correlation coefficients cannot be calculated; thus, distance correlation coefficients are used instead. This method not only measures linear correlations but also captures non-linear correlations \cite{distance_correlation}. The resulting matrices $C_{ef}\in \mathbb{R}^{30\times 36}$ and $C_{fe}\in \mathbb{R}^{36\times 30}$ are transposes of each other. To enable 2-dimensional convolutions, the matrices' last dimension is projected onto a $16\times 16$ plane, constructed based on the distribution of $66$ scalp channels (see Fig. \ref{correlation}). On this plane, each point is either set to $0$ or represents a correlation coefficient value. Thus, $C_{ef}$ and $C_{fe}$ are transformed into shapes of $30\times 16\times 16$ and $36\times 16\times 16$, respectively. Each plane can be viewed as the spatial distribution of channel correlations.

Next, a 2-dimensional convolution with $d_{in}$ kernels is applied to $C_{ef}$, yielding the query $Q\in \mathbb{R}^{d_{in}\times 16\times 16}$, where $d_{in}$ is the number of input EEG representation channels. $Q$ selects $d_{in}$ features for representation mapping from the EEG data. A similar operation on $C_{fe}$ results in the key $K\in \mathbb{R}^{d_{out}\times 16\times 16}$, where $d_{out}$ is the number of output fNIRS representation channels. $K$ determines $d_{out}$ relevant fNIRS features. The EEG representation serves as the value $V$ in the attention mechanism. The output of the SCG module is formulated as follows:
\begin{align}
output =Score \cdot V=Softmax\big(\dfrac{QK^{T}}{\sqrt{d_{out}}}\big) \cdot V
\end{align}
where $Score\in \mathbb{R}^{d_{in}\times d_{out}}$ is the attention score matrix, with each element representing the contribution weight of a particular EEG channel to a specific fNIRS channel. Since EEG channels will assign higher weights to spatially adjacent fNIRS channels, the weight information obtained through 2-dimensional convolution more precisely reflects the spatial correspondence between EEG and fNIRS channels.

In summary, the SCG module employs an attention mechanism based on inter-channel correlation and enhances the traditional 1-dimensional attention mechanism by executing 2-dimensional convolution operations to obtain $Q$ and $K$. The inter-channel correlation guides the representation mapping. Although this relationship is based on statistical rather than spatio-temporal features, unified representation learning of the EEG and fNIRS is performed before mapping. Additionally, the 2-dimensional convolution preserves the spatial distribution features of the correlations, maintaining the spatial relationship between EEG and fNIRS channels. Therefore, this mapping can adaptively leverage the self-attention mechanism. This mechanism enables the attention module to focus more on EEG spatio-temporal features that are more relevant to fNIRS while ignoring irrelevant features, thus achieving precise mapping from EEG representation to fNIRS representation.

\subsubsection{Spatial Representation}
The input may include only $e_0$ and the EEG representation. In such instances, the SCG module learns the spatial representation of EEG. The method for obtaining the correlation coefficient matrix $C_e$ from $e_0$ is similar to that of $C_{ef}$ and $C_{fe}$ but simpler. There are two main differences: first, the correlation is measured by the general Pearson correlation coefficient; second, two identical $C_e$ matrices are used in different convolutions to calculate $Q$ and $K$. Similarly, if the input consists of $f_0$ and the fNIRS representation, the output will also be the fNIRS representation.

2-dimensional convolution has local perception capabilities, enabling it to effectively capture the spatial structure information of the data. This enhances integration between adjacent channels and mitigates blind fusion between channels seen in 1-dimensional convolution. Consequently, the correlations obtained via 2-dimensional convolution reflect the spatial relationships between channels. An attention mechanism based on inter-channel correlations can amplify the contributions of relevant channels while excluding the interference of irrelevant ones, thereby enhancing the critical spatial information of the data.

\subsection{MTR Module}

The MTR module halves or doubles the length of the input sequence without changing the number of channels.

The multi-scale 1-dimensional depth-wise convolution comprises four convolutions with kernel sizes of 3, 5, 7, and 9, each with a stride of 1. Each kernel individually performs convolutions on a single channel, enabling the network to focus on specific temporal feature distributions without inter-channel interference. Varying kernel sizes enable multi-scale temporal feature extraction, effectively capturing dynamic brain activity changes across various frequencies and time spans. The causal dilated convolution is implemented with three consecutive convolutions with dilation rates of 1, 2, and 4, a stride of 2, and a kernel size of 2, with zero-padding added to the left side. This convolutional structure expands each output point's receptive field and makes it depend solely on its predecessors, thus ensuring high precision and causality in the extracted temporal features.

For downsampling, the 1-dimensional point-wise convolution involves four convolutions with a kernel size of 1 and a stride of 2. Two convolutions use no padding, while the other two use zero-padding on the left side to include all sequence points in the convolution operation. Point-wise convolutions help integrate information from different channels. Using four, rather than just one, point-wise convolutions allows for adequate learning of the latent temporal relationships between channels. For upsampling, two transposed convolutions and two bilinear interpolations with a scaling factor of 2 are employed, effectively avoiding the potential information loss that might occur with a single method.

Each convolution in the MTR module is followed by an activation function that transforms the linear operations into nonlinear ones, enhancing the model's expressive capability.

Overall, the combination of depth-wise separable convolutions and causal dilated convolutions enables the MTR module to precisely capture the brain activity features in EEG and the hemodynamic changes in fNIRS, establishing a robust foundation for EEG-to-fNIRS representation mapping. These techniques enhance the temporal feature extraction capabilities of the model and improve the temporal consistency and accuracy of feature representation in cross-modal generation.

\section{Dataset}

The study used a publicly available dataset of EEG and fNIRS recordings from 29 healthy subjects performing left and right-hand MI tasks \cite{dataset}. The experimental procedure comprised three separate sessions, each with 20 trials. During each trial, subjects were instructed to imagine executing a grasping movement with either their left or right hand, as indicated by the direction of a black arrow displayed at the center of a screen for 2 seconds. Following the trial initiation, subjects engaged in imagining the grasping movement for 10 seconds until hearing a brief beep and seeing the word "STOP" on the screen, signifying the trial's end. Rest periods of 14 to 16 seconds were interleaved between each trial.

The EEG data were recorded from 30 active electrodes positioned according to the international 10-5 system, resulting in 30 channels sampled at 1000Hz. The data were downsampled to 200Hz by the provider. fNIRS signals were recorded using 14 sources and 16 detectors placed over the frontal, motor, and visual areas, totaling 36 channels sampled at 12.5Hz and later downsampled to 10Hz. The sensors' spatial arrangement and recording site distribution of EEG-fNIRS can be referenced in Fig. \ref{correlation}.

In this experiment, the EEG signals were re-referenced to a common average reference and filtered using a 4th order Chebyshev Type II bandpass filter with a passband of 0.5 to 50 Hz, then downsampled to 160Hz. Ocular artifacts were removed via independent component analysis. The fNIRS data were filtered using a 6th order zero-phase Butterworth bandpass filter with a passband of 0.01 to 0.1 Hz after conversion from raw recordings to deoxy-hemoglobin (HbR) and oxy-hemoglobin (HbO) data. HbR and HbO, collectively referred to as fNIRS, were uniformly processed throughout the experiment. Finally, the three datasets, EEG, HbR, and HbO, were segmented into epochs. Each dataset consisted of 1740 samples, representing the 1740 MI trials across all 29 subjects. The time series length for EEG recordings was 4000, corresponding to a duration of 25 seconds, while for fNIRS recordings, it was 256, corresponding to 25.6 seconds. Each trial included a 5-second preparation period, followed by a 10-second trial period, and then a rest period.

\section{Result}

\subsection{Classification}

\begin{figure}[htbp]
	\centering
	\begin{minipage}{\linewidth}
	    \centering
	    \centerline{(a) EEG-HbR Group}\medskip
		\includegraphics[width=\linewidth]{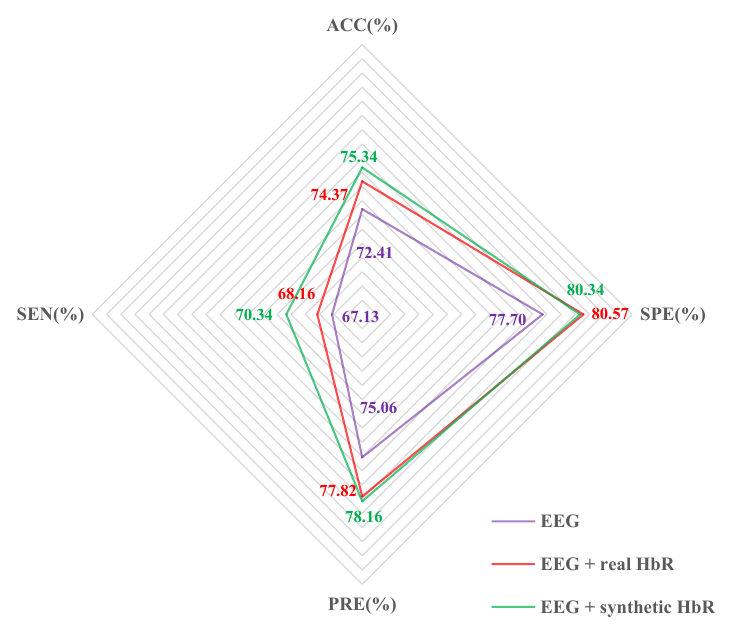}
		\label{hbr_radar}
	\end{minipage}
	\vfill
	\begin{minipage}{\linewidth}
		\centering
		\centerline{(b) EEG-HbO Group}\medskip
		\includegraphics[width=\linewidth]{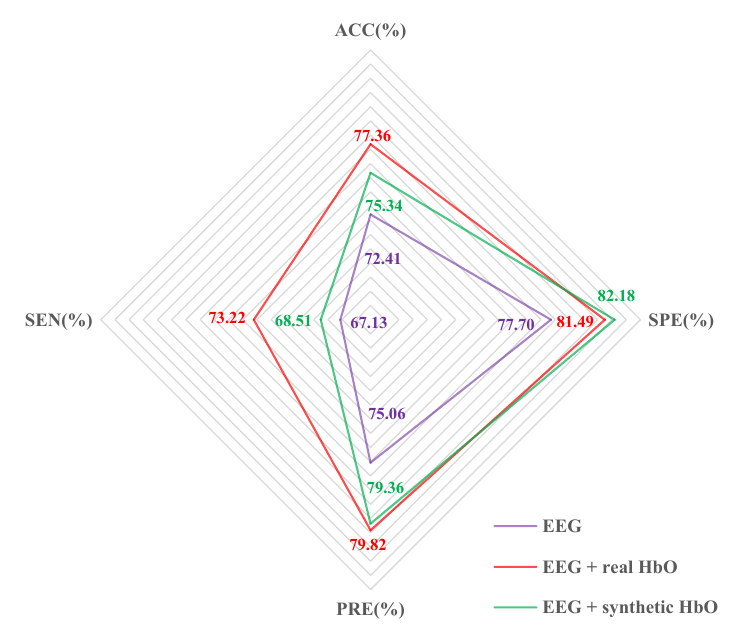}
		\label{hbo_radar}
	\end{minipage}
\caption{Radar chart of the classification results.}
\label{radar}
\end{figure}

This study investigates the potential of synthetic fNIRS to substitute real fNIRS when combined with EEG for left/right motor imagery (LMI/RMI) classification. The classification results of five signal combinations were compared: EEG, EEG-real HbR, EEG-synthetic HbR, EEG-real HbO, and EEG-synthetic HbO. The classification was implemented using models proposed in \cite{eeg_fnirs_2}, where ESNet served as the classifier for single EEG signals and FGANet for EEG-fNIRS signals. In MI-BCI applications, LMI and RMI sample distributions may be imbalanced, so the study evaluated not only accuracy (ACC) but also sensitivity (SEN), specificity (SPE), and precision (PRE) to comprehensively compare performance. ACC assessed overall classification accuracy, PRE evaluated LMI classification accuracy (the proportion of correctly identified LMI samples among those classified as LMI), SEN measured the ability to identify LMI (the proportion of correctly identified LMI samples), and SPE measured the ability to identify RMI. The classifier was trained and tested under different sample label ratios of 2:8, 3:7, 4:6, 5:5, 6:4, 7:3, and 8:2, averaging five test results to obtain the final outcomes. The results were divided into EEG-HbR and EEG-HbO groups, as depicted in Fig. \ref{radar}.

All metrics for the hybrid signals significantly outperformed EEG, consistent with previous findings \cite{eeg_fnirs_2}, affirming classifier reliability and the necessity of EEG-fNIRS cross-modal generation. Despite slightly lower SPE than EEG-real HbR, EEG-synthetic HbR showed higher values in the other three metrics, indicating only a minor setback in LMI recognition. However, this discrepancy was acceptable, as the difference in correctly predicted RMI sample numbers was only 2. Apart from SPE, EEG-synthetic HbO displayed lower metrics compared to EEG-real HbO. It is due to higher correctly predicted RMI samples in EEG-synthetic HbO but significantly lower correctly predicted LMI samples (a gap of 41 compared to EEG-real HbO), thereby substantially reducing SEN and lowering the other three metrics. The visualization of hemodynamic responses provides a credible explanation for these outcomes. Despite some differences, synthetic HbO signals perform comparably to real HbO in joint classification with EEG, indicating their potential substitutability.

\subsection{Spatial correspondence with EEG}

\begin{figure*}[!t]
	\centering
	\includegraphics[width=\linewidth]{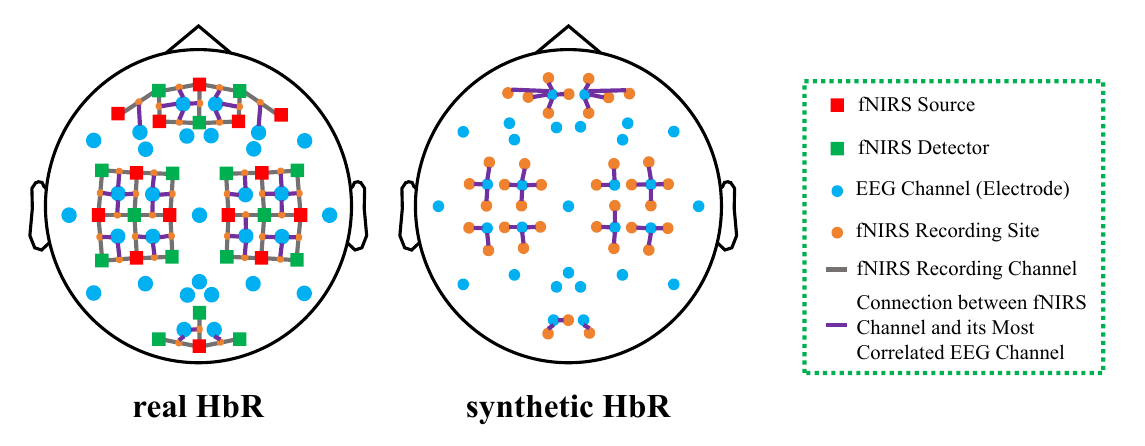}
    \caption{The distribution of the most correlated EEG channels with real and synthetic HbR signals. The section concerning real HbR also illustrates the sensors' spatial arrangement and recording site distribution of EEG-fNIRS.}
    \label{correlation}
\end{figure*}

The position of each fNIRS channel is considered the center of the fNIRS source and detector and is expected to be most closely correlated with the nearest EEG channel. The study investigates whether this spatial relationship between real fNIRS recordings and EEG recordings is preserved in synthetic fNIRS signals. Distance correlation coefficients were used to quantify these correlations. For each fNIRS channel, the EEG channel with the highest correlation coefficient was selected as the most correlated EEG channel. Fig. \ref{correlation} illustrates the distribution of the most correlated EEG channels for real HbR and synthetic HbR. While some synthetic channels differed from their corresponding real channels in EEG selection, both choices were within the same brain region, indicating that synthetic HbR and EEG achieved at least a region-level correspondence. Additionally, each HbR channel that selected an incorrect EEG channel was actually adjacent to two EEG channels, making it difficult to determine which of the two was spatially closest to this HbR channel. Therefore, it is reasonable to select either of the two as the most correlated EEG channel. The distribution of the most correlated EEG channels for real HbO and synthetic HbO signals was entirely consistent, so the results are not shown. In summary, the experimental results demonstrate that synthetic fNIRS signals retain spatial correspondence with EEG signals.

\subsection{Hemodynamic Response Curve}

\begin{figure*}[!t]
    \centering
    \includegraphics[width=\linewidth]{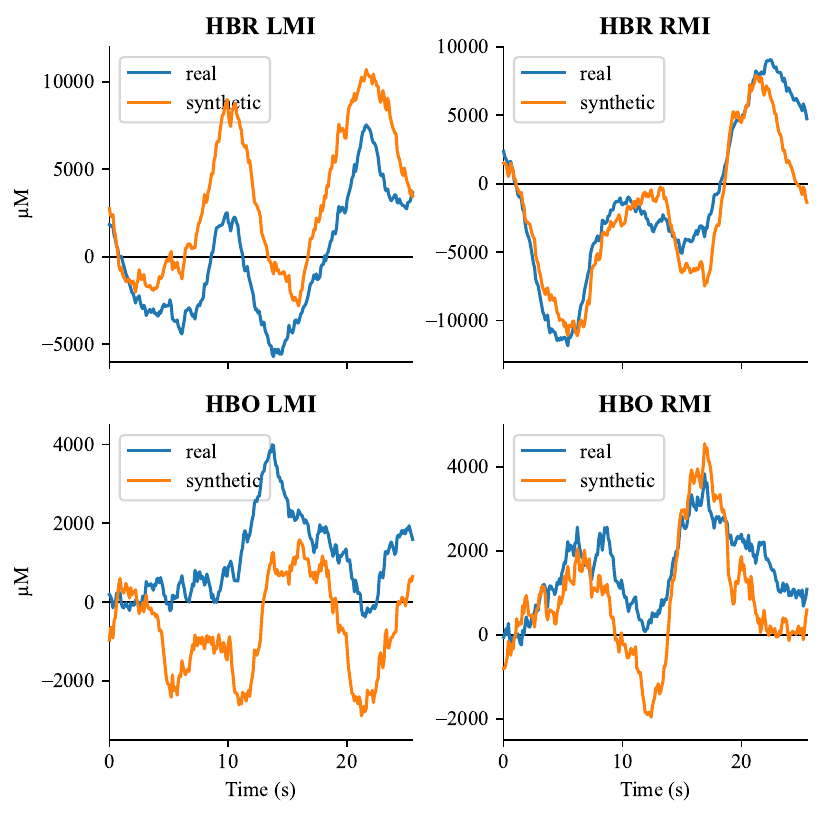}
    \caption{Comparison of hemodynamic response curves between real and synthetic HbR/ HbO under LMI and RMI tasks.}
    \label{evokeds}
\end{figure*}

This study compares the hemodynamic response curves of HbR and HbO signals under LMI and RMI tasks by averaging epochs across all subjects, as shown in Fig. \ref{evokeds}. These curves visually represent the temporal characteristics of fNIRS signals.

The synthetic and real HbR curves show similar trends during the LMI task, despite significant numerical differences between 5 and 15 seconds. In the RMI task, both curves are closely aligned in trend and value, particularly within the initial 15 seconds. For synthetic and real HbO curves, substantial differences are observed during the LMI task, but the trend differences diminish after 15 seconds; conversely, they are quite similar during the RMI task. Overall, the hemodynamic response curves of synthetic and real fNIRS signals are generally consistent, reflecting similar temporal characteristics.

The averaged hemodynamic response curve can explain the classification results to some extent. In the LMI task, the synthetic HbR curve is globally amplified compared to the real HbR curve, thereby magnifying the differences between synthetic HbR curves in LMI and RMI tasks. This amplification indicates that all epoch curves contributing to the synthetic HbR curve collectively enhance the distinctions between LMI and RMI tasks. This effect likely aids classifiers in identifying potential features of synthetic HbR signals' hemodynamic responses across both tasks. The superior classification performance of EEG-synthetic HbR compared to EEG-real HbR provides credible evidence for this hypothesis. However, the synthetic HbO curve during the LMI task appears more complex than the real HbO curve, indicating great variability among epoch curves, potentially reducing the distinguishability of the two curves under LMI and RMI tasks and thereby weakening the classification ability of EEG-synthetic HbO for LMI.

\subsection{Scalp Topography}

\begin{figure*}[htbp]
	\centering
	\begin{minipage}{\linewidth}
	    \centering
	    \centerline{(a) HbR LMI}\medskip
		\includegraphics[width=0.9\linewidth]{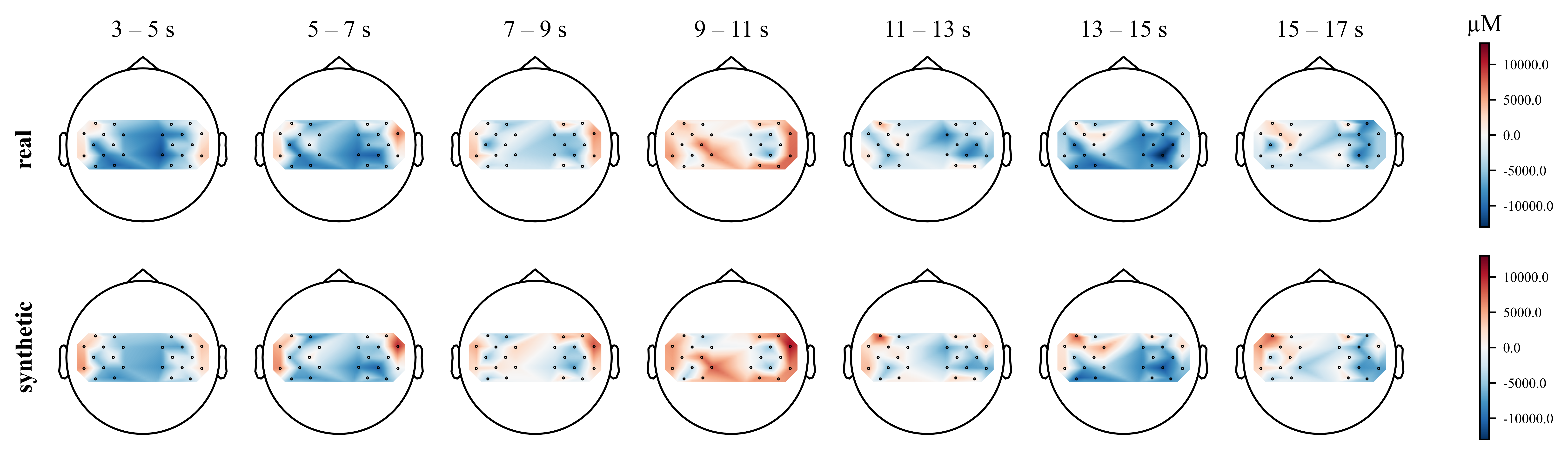}
		
		\label{tp_hbr_lmi}
	\end{minipage}
	\vfill
	\vspace{10pt}
	\begin{minipage}{\linewidth}
		\centering
		\centerline{(b) HbR RMI}\medskip
		\includegraphics[width=0.9\linewidth]{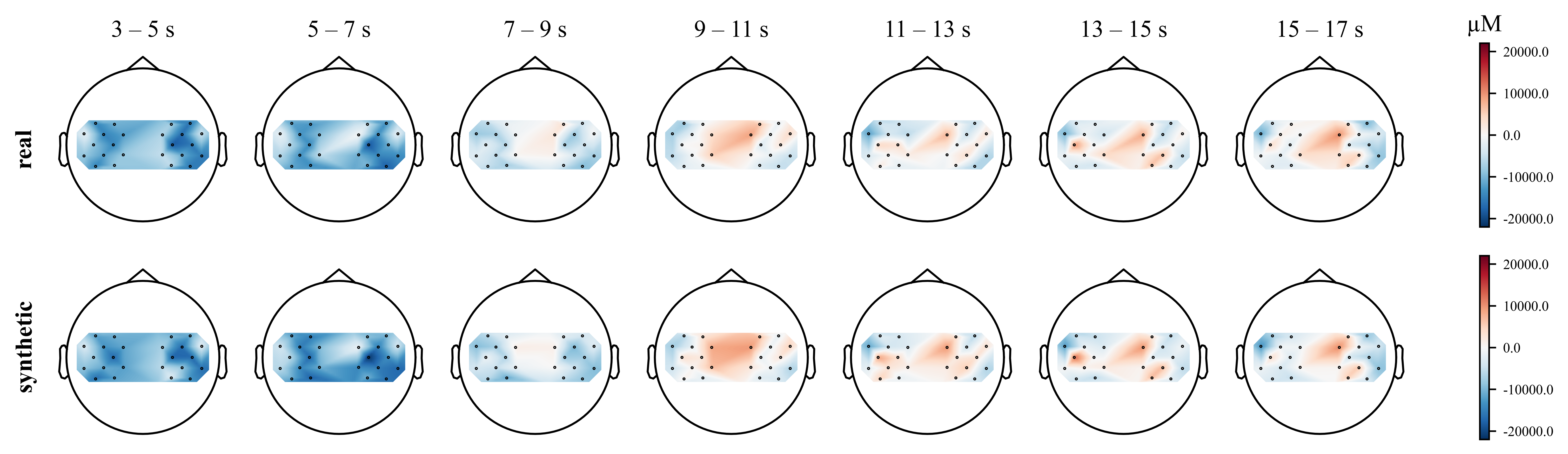}
		\label{tp_hbr_rmi}
	\end{minipage}
	\vfill
	\vspace{10pt}
	\begin{minipage}{\linewidth}
		\centering
		\centerline{(c) HbO LMI}\medskip
	
		\includegraphics[width=0.9\linewidth]{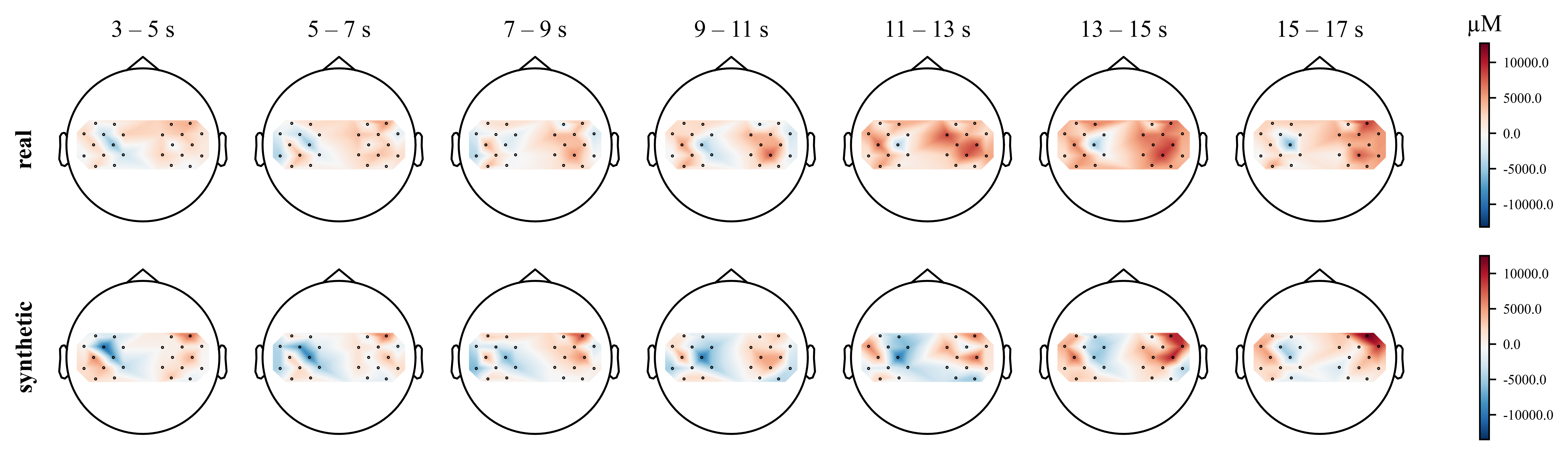}
		
		\label{tp_hbo_lmi}
	\end{minipage}
	\vfill
	\vspace{10pt}
	\begin{minipage}{\linewidth}
		\centering
		\centerline{(d) HbO RMI}\medskip
		\includegraphics[width=0.9\linewidth]{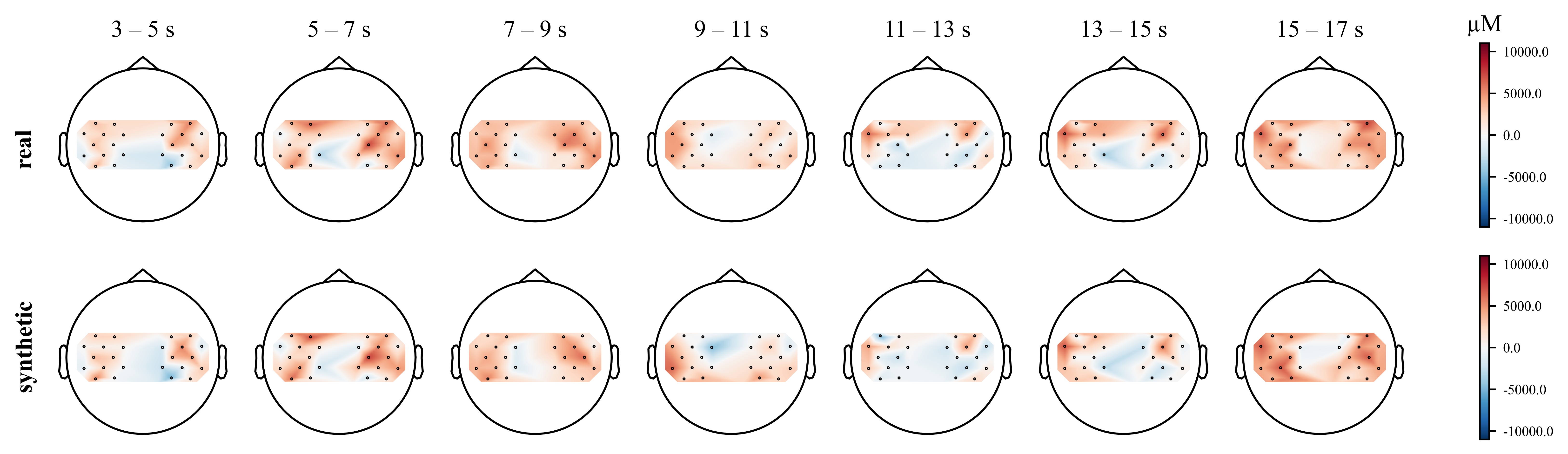}
	    
		\label{tp_hbo_rmi}
	\end{minipage}
\caption{Comparison of scalp topography between real and synthetic HbR/ HbO under LMI and RMI tasks.}
\label{topography}
\end{figure*}

The scalp topography of fNIRS illustrates the concentration distribution of deoxyhemoglobin or oxyhemoglobin across different regions of the brain, which can characterize the spatial features of fNIRS signals. This study qualitatively analyzes the spatial distribution differences between synthetic and real fNIRS over a time period of 3 to 17 seconds by comparing seven scalp topographies, as shown in Fig. \ref{topography}. For clarity, only channels from the motor area are displayed, and linear interpolation is applied to calculate points outside the channel locations.

Under the LMI task, the spatial distributions of synthetic and real HbR are generally consistent across all seven time periods, with a few channels exhibiting noticeable differences at certain time points. For the RMI task, the spatial distributions are consistent. For synthetic and real HbO under the LMI task, there are notable differences in some channels for each time period. In the RMI task, the distributions are similar across all time periods, though there are differences in magnitude, particularly during the 11 to 13 second period. In summary, the spatial distributions of real and synthetic signals are more consistent in the RMI task, whereas in the LMI task, they demonstrate greater complexity and variability. This could be one reason why classifiers find it easier to identify RMI task features. The visualization results indicate a relatively high degree of spatial similarity between synthetic and real signals.

\section{Ablation Study}

\begin{table*}[!ht]
\centering
    \caption{Classification results of ablation experiments}
    \label{ablation}
    \begin{tabular}{c c|cccc}
        \hline
        Modality & Modules & ACC(\%) & SPE(\%) & PRE(\%) & SEN(\%) \\
        \hline
        \multirow{6}{*}{HbR} & ATTN + COV & 68.56 & 70.23 & 69.20 & 66.90 \\

        & ATTN + MTR &  65.06 & 66.32 & 65.45 & 63.79 \\

        & SCG(EEG) + COV & 70.46 & 71.61 & 70.94 & 69.31 \\

        & SCG(EEG) + MTR & 75.34 & 80.34 & 78.16 & 70.34 \\
  
        & SCG(fNIRS) + COV & 72.93 & 76.44 & 74.66 & 69.43\\
   
        & SCG(fNIRS) + MTR & \textbf{76.72} & \textbf{80.69} & \textbf{79.03} & \textbf{72.76} \\
        \hline
        \multirow{6}{*}{HbO} & ATTN + COV &  67.18 & 67.82 & 67.40 & 66.55 \\
        & ATTN + MTR & 66.09 & 66.78 & 66.32 & 65.40 \\
        & SCG(EEG) + COV & 71.38 & 76.44 & 73.79 & 66.32 \\
        & SCG(EEG) + MTR & 75.34 & 82.18 & 79.36 & 68.51 \\
        & SCG(fNIRS) + COV & 73.33 & 78.85 & 76.23 & 67.82 \\
        & SCG(fNIRS) + MTR & \textbf{77.41} & \textbf{82.87} & \textbf{80.77} & \textbf{71.95} \\
        \hline
    \end{tabular}
\end{table*}

This study conducted five ablation experiments to evaluate the contributions of the SCG and MTR modules to SCDM performance. Five module combinations were utilized: ATTN + COV, SCG (fNIRS) + COV, SCG (EEG) + COV, ATTN + MTR, and SCG (fNIRS) + MTR, generating synthetic fNIRS signals. EEG-synthetic fNIRS was then used in classification experiments to evaluate each combination's performance. The 1-dimensional multi-head self-attention mechanism (ATTN) in the denoising diffusion probabilistic model (DDPM) served as the control for the SCG module. In line with DDPM's U-Net architecture, the MTR module's depthwise separable and causal dilation convolutions were replaced with conventional convolutions, forming the COV module, which acted as the MTR module's control. The SCG module has two forms, SCG (fNIRS) and SCG (EEG), sharing the same architecture. SCG (fNIRS) only accepts fNIRS sequences as input to learn their spatial representations. Since the proposed SCDM adopts the SCG (EEG) + MTR combination, the SCG (fNIRS) + MTR combination can be viewed as a variant of SCDM that enhances fNIRS rather than achieving cross-modal generation from EEG to fNIRS. ATTN + COV represents a variant of DDPM and serves as the baseline for SCDM.

The ablation experiment results are presented in Table \ref{ablation}. Consistent patterns were observed for both the HbR and HbO groups:

(1) The MTR module enhances performance only when combined with the SCG module; its performance is inferior to the COV module when used with the Attn module.

(2) The SCG module significantly outperforms the ATTN module. The classification performance of the ATTN + COV and ATTN + MTR combinations is markedly inferior to that of the other four combinations. This underscores the SCG module's critical role in spatial information extraction and cross-modal generation, highlighting the proposed SCDM's superiority over DDPM.

(3) The SCG (fNIRS) module outperforms the SCG (EEG) module. This can be attributed to unimodal SCDM's easier generation of synthetic sequences consistent with the real fNIRS distribution compared to cross-modal SCDM. However, minor differences across the four classification metrics indicate that the proposed cross-modal SCDM approaches the performance of unimodal SCDM.

In summary, the contributions of the two core modules to SCDM are substantial. The SCG module is crucial for cross-modal generation, and the MTR module significantly enhances SCDM performance when combined with the SCG module.

\section{Conclusion}

The study proposes the SCDM for cross-modal generation of synthetic fNIRS signals from EEG data. Synthetic fNIRS has relatively high substitutability for real fNIRS, suggesting that the cross-modal generation approach can serve as a new paradigm for acquiring hybrid EEG-fNIRS signals in MI-BCIs. Ablation studies validate the crucial role of the SCG module in cross-modal generation and demonstrate that the MTR module significantly complements the SCG module, collectively enhancing the performance of the SCDM framework.

However, this study warrants further exploration. Firstly, ablation experiments specifically targeting the SCG module are necessary. Although the SCG module has dual functions of spatial representation and representation mapping, the individual contributions of these functions to the model remain unclear. Furthermore, there are differences between synthetic fNIRS signals and real fNIRS signals in terms of the hemodynamic response curve and scalp topography, which may indicate limitations in the representation mapping capability of the SCG module. Secondly, the SCDM should be applied to various MI datasets to assess its generalization capability, and it should even be tested on datasets from other tasks, such as mental arithmetic, to broaden the application of EEG-fNIRS cross-modal generation in BCI. Finally, given that representation learning involves both temporal and spatial dimensions, 3-dimensional convolutions suitable for spatio-temporal data \cite{Cov3D_1, Cov3D_2} can be considered to enhance SCDM algorithmically.

\section*{Acknowledgment}

\bibliographystyle{ieeetr}
\bibliography{reference}

\end{document}